\documentclass[11pt]{article}
\usepackage[utf8]{inputenc}
\usepackage[T1]{fontenc}
\usepackage{lmodern}
\usepackage[a4paper,margin=1in]{geometry}
\usepackage{amsmath,amssymb}
\usepackage{graphicx}
\usepackage{booktabs,longtable,array,multirow,calc}
\usepackage{arydshln}
\usepackage{caption}
\usepackage{float}
\usepackage[numbers,super,sort&compress]{natbib}
\usepackage{hyperref}
\hypersetup{colorlinks=true,linkcolor=blue,citecolor=blue,urlcolor=blue}
\graphicspath{{figures/}}
\DeclareUnicodeCharacter{03C0}{\ensuremath{\pi}}
\DeclareUnicodeCharacter{0393}{\ensuremath{\Gamma}}
\DeclareUnicodeCharacter{0394}{\ensuremath{\Delta}}
\DeclareUnicodeCharacter{03BB}{\ensuremath{\lambda}}
\DeclareUnicodeCharacter{03BC}{\ensuremath{\mu}}
\DeclareUnicodeCharacter{2248}{\ensuremath{\approx}}
\DeclareUnicodeCharacter{00B1}{\ensuremath{\pm}}
\DeclareUnicodeCharacter{00D7}{\ensuremath{\times}}
\DeclareUnicodeCharacter{03B3}{\ensuremath{\gamma}}
\DeclareUnicodeCharacter{2212}{-}
\DeclareUnicodeCharacter{2013}{--}
\DeclareUnicodeCharacter{2014}{---}

\title{Non-volatile integrated photonics on lithium tantalate-on-insulator}
\author{%
Yuhang Li\textsuperscript{1,*},
Miao Deng\textsuperscript{1,*},
Xun Zhang\textsuperscript{2,*},
Cheng Zeng\textsuperscript{1,\textdagger},
Chijun Li\textsuperscript{1},
Yiqi Dai\textsuperscript{1},\\[0.2em]
Yuankang Huang\textsuperscript{1},
Siyu Lu\textsuperscript{1},
Zhenwu Mo\textsuperscript{1},
Xiao Wu\textsuperscript{1},
Peng Tan\textsuperscript{1,3}\\[0.2em]
Yong Zhang\textsuperscript{2},
Yikai Su\textsuperscript{2},
Jinsong Xia\textsuperscript{1,\textdagger}\\[0.8em]
\small \textsuperscript{1}Wuhan National Laboratory for Optoelectronics and School of Optical and Electronic Information,\\
Huazhong University of Science and Technology, Wuhan 430074, China.\\
\small \textsuperscript{2}State Key Laboratory of Photonics and Communications, Department of Electronic Engineering,\\
Shanghai Jiao Tong University, Shanghai, China.\\
\small \textsuperscript{3}School of Physics, Harbin Institute of Technology, Harbin 150001, China.\\[0.5em]
\small \textsuperscript{*}These authors contributed equally: Yuhang Li, Miao Deng and Xun Zhang.\\
\small \textsuperscript{\textdagger}Correspondence to Cheng Zeng
(\href{mailto:zengchengwuli@hust.edu.cn}{zengchengwuli@hust.edu.cn}) or\\
\small Jinsong Xia
(\href{mailto:jsxia@hust.edu.cn}{jsxia@hust.edu.cn}).
}
\date{}

\begin{document}
\maketitle

\begin{abstract}

Scalable reconfigurable photonic integrated circuits require low-loss, high-speed optical control
without continuous holding power. Yet widely used thermo-optic tuning and continuously biased
electro-optic tuning consume static power and introduce thermal crosstalk or bias drift. Here we
demonstrate a monolithic non-volatile photonics platform on lithium tantalate-on-insulator (LTOI).
In congruent x-cut lithium tantalate, the switched ferroelectric-domain configuration is retained
after the write field is removed. On the same LTOI platform, we demonstrate a waveguide
propagation loss of approximately 0.05--0.06 dB/cm and multilevel non-volatile phase tuning in
separate devices. The programmed states remain distinguishable through \(10^{6}\) write cycles.
Weighted segmented electrodes resolve 137 phase
positions across a \(\pi\) range, corresponding to an analogue phase-setting resolution of
\(\sim0.007\pi\). We further combine non-volatile phase control with high-speed electro-optic modulation to
achieve zero-static-power bias control of a >110 GHz modulator and a 59.3-dB extinction ratio after
non-volatile trimming. At the system level, an image-edge-detection chip achieves a measured energy
efficiency of \(3.48~\mathrm{TOPS\,W^{-1}}\). These
results establish LTOI as an integrated photonics platform that combines persistent optical
reconfigurability with low-loss routing and high-speed electro-optic modulation.

\end{abstract}

\section{Introduction}
\label{introduction}

Photonic integrated circuits (PICs) are evolving from isolated high-performance devices into
large-scale reconfigurable systems that combine optical routing, filtering, modulation and signal
processing. These systems increasingly support optical communications, photonic computing, microwave
photonics and quantum technologies.\cite{Bogaerts2020Programmable} Their scalability is determined not by any single device metric,
but by the simultaneous availability of low optical loss, high-speed drive capability and low static
control power. Low loss preserves optical power margins across cascaded devices and interferometric
networks. High-speed drive capability governs how effectively photonic chips interface with high-speed
electronics, optical communication links and microwave-photonic processing systems. Low static control
power becomes critical as the number of tunable elements increases, because continuously biased phase
shifters and resonators accumulate power consumption, thermal crosstalk and feedback overhead.
\cite{PerezLopez2020SelfConfiguration,Padmaraju2012Thermal} These requirements are particularly acute in large high-speed modulator arrays, where each channel needs a
stable operating point.\cite{Padmaraju2012Thermal} They also make post-fabrication trimming essential in large-scale photonic
systems, where small process variations accumulate across many components and degrade programmed
transfer functions.\cite{Hamerly2022FaultTolerant,Jayatilleka2021PostFabrication} These constraints also appear in edge-intelligence processors. Image edge detection
is one representative primitive: it extracts compact spatial features near the sensor and reduces data
movement and latency in autonomous systems, robotics, industrial inspection and remote sensing.
\cite{Bai2023Microcomb} Beyond this example, many photonic applications must sustain high-speed drive while operating within tight
energy and control budgets. Low optical loss, high-speed drive capability and low static control power
are therefore key requirements for next-generation scalable integrated photonics platforms.

In present PICs, programmable optical states are usually defined by thermo-optic actuators or
electro-optic electrodes held under a continuous bias.\cite{PerezLopez2020SelfConfiguration,Padmaraju2012Thermal} These elements can compensate fabrication
variations and set the operating points of Mach-Zehnder interferometer (MZI) switch arrays,
microring resonators, arrayed waveguide gratings and high-speed modulators. Their programmed
states, however, are volatile. Maintaining a phase shift, resonance wavelength, splitting ratio or
modulator bias point requires continuous electrical power.\cite{PerezLopez2020SelfConfiguration,Padmaraju2012Thermal} As photonic systems grow, the associated
static power, thermal crosstalk, bias drift and feedback-control overhead become increasingly
difficult to manage.\cite{PerezLopez2020SelfConfiguration} For universal \(N\)-mode interferometer meshes,
\(N(N-1)/2\) tunable two-mode elements are required, and the number of controls therefore scales as
\(O(N^2)\).\cite{Clements2016Optimal} Representative silicon implementations report thermo-optic
\(P_{\pi}\) values from \(1.7~\mathrm{mW}\) to approximately \(10~\mathrm{mW}\).
\cite{Miller2020LargeScale,Annoni2017Unscrambling} The static power of a large mesh or modulator
array can readily reach the watt scale.\cite{Miller2020LargeScale} The resulting power density demands active thermal
management and limits dense PICs.\cite{PerezLopez2020SelfConfiguration,Padmaraju2012Thermal} These constraints motivate non-volatile integrated photonic
platforms.

Non-volatile photonics seeks to retain programmed optical states after the write stimulus is
removed. Several integrated approaches have been explored, but each leaves a gap between low optical
loss, high-speed electro-optic operation and scalable integration. Phase-change materials (PCMs)
can store optical states through thermally written structural phases.\cite{Wuttig2017PCM} An
\(\mathrm{Sb_2Se_3}\)-based platform provides a large non-volatile refractive-index contrast
(\(\Delta n = 0.77\)).\cite{Yang2023NonVolatilePCM} However, the thermally written material states introduce appreciable loss
(\(\sim200~\mathrm{dB\,cm^{-1}}\)) and potential thermal crosstalk; more than \(10^{4}\) reversible
switching cycles have been demonstrated.\cite{Wuttig2017PCM,Yang2023NonVolatilePCM} Integrated microelectromechanical systems (MEMS) switches can latch
waveguide positions and achieve an extinction ratio of \(44.4~\mathrm{dB}\).\cite{Hu2025NonvolatileMEMS} They generally require
suspended mechanical structures, and the reported non-volatile latching device provides digital
two-state rather than multilevel tuning.\cite{Hu2025NonvolatileMEMS} These PCM and MEMS
devices are also commonly implemented on silicon-on-insulator (SOI) platforms, where high-speed
electro-optic modulation beyond \(110~\mathrm{GHz}\) remains challenging. Beyond PCM and MEMS, ferroelectric non-volatile photonics has been explored in
barium titanate (BTO) and lead zirconate titanate (PZT).\cite{GelerKremer2022BTO,Li2025PZTMemristor} In these platforms,
ferroelectric domain switching stores the optical state, while the Pockels effect
can support high-speed electro-optic modulation.\cite{GelerKremer2022BTO,Li2025PZTMemristor} BTO phase shifters have achieved
non-volatile operation using \(\pm5~\mathrm{V}\) write-pulse trains with a total programming duration
of approximately \(60~\mu\mathrm{s}\).\cite{CatalaLahoz2026BTOFPPGA} However, reported non-volatile BTO devices have not
yet combined this functionality with ultralow-loss waveguides. A representative
multilevel BTO phase shifter showed a propagation loss of
\(4.8 \pm 0.2~\mathrm{dB\,cm^{-1}}\).\cite{GelerKremer2022BTO} In a recent BTO programmable circuit
heterogeneously integrated with SOI photonics, the programmable unit cell had an insertion loss of
\(1.48~\mathrm{dB}\), attributed mainly to interlayer mode conversion.
\cite{CatalaLahoz2026BTOFPPGA} BTO also has a large and dispersive dielectric permittivity,
which increases microwave loading and makes impedance matching and
microwave--optical velocity matching difficult.\cite{Chelladurai2025BTOPermittivity} PZT provides a strong Pockels
response and non-volatile domain switching, but its high dielectric permittivity
can introduce similar RF constraints.\cite{Li2025PZTMemristor} These factors make it difficult to combine
low-voltage operation with ultrabroadband electro-optic modulation in BTO and PZT
photonic platforms. Taken together, current approaches leave a clear platform gap:
persistent optical control, low-loss routing and high-speed electro-optic
operation have not yet been demonstrated together on a single platform.
\cite{GelerKremer2022BTO,CatalaLahoz2026BTOFPPGA,Li2025PZTMemristor}

Thin-film lithium tantalate (LT) is emerging as a new integrated photonics platform with capabilities
similar to those of thin-film lithium niobate (LN). It offers a comparable Pockels coefficient and has already supported
low-loss waveguides and high-bandwidth electro-optic modulators.\cite{Wang2024Nature,Wang2024Optica} Relative to LN, LT provides a
higher photorefractive-damage threshold, lower birefringence and a more stable low-frequency
electro-optic response.\cite{Wang2024Nature,Powell2024DCStable,Wang2025HighPower} However, phase control in lithium tantalate-on-insulator (LTOI) and lithium
niobate-on-insulator (LNOI) waveguides still relies mainly on thermo-optic tuning or static
electro-optic biasing.\cite{Wang2024Nature,Bente2023PatternGeneration} Non-volatile phase tuning generally requires heterogeneous integration of
phase-change materials.\cite{Bente2023PatternGeneration,Xu2025PockelsMemory} Defect models for congruent LT motivate testing a possible route in
integrated waveguides: whether controlled ferroelectric domain inversion can provide intrinsic
non-volatile phase control in LTOI.\cite{Gopalan1996InternalField,Kim2001DomainReversal} LT therefore offers a promising single-material
platform in which non-volatile phase control can coexist with low-loss routing and high-speed
electro-optic modulation.

Here we demonstrate, to our knowledge, the first monolithic LTOI platform for intrinsic,
ferroelectric-domain-based non-volatile phase control without an added state-retentive material. The
platform combines low optical loss (\(\sim0.05~\mathrm{dB\,cm^{-1}}\)) and high electro-optic
modulation bandwidth (\(>110~\mathrm{GHz}\)). We first translate the defect-dipole model of
congruent LT into integrated waveguides. In this model, the defect dipoles generate a built-in field
in the initial state. Ferroelectric-domain reversal changes the effective electro-optic coefficient.
Because lithium vacancies have limited mobility at room temperature, most defect dipoles retain
their initial orientation during domain reversal. The built-in field therefore retains its original
direction and most of its magnitude after switching. Its interaction with the modified effective
electro-optic coefficient generates a persistent refractive-index change through the Pockels effect. Based on
this mechanism, we develop a stable and repeatable method for non-volatile phase control in LT. The
method uses local domain programming in monolithic LTOI waveguides and is
compatible with existing thin-film LT photonic fabrication flows. We validate it in asymmetric MZIs
and microring resonators, demonstrating multilevel tuning, high phase-setting precision, long-term
retention, repeated-programming endurance and essentially unchanged microring Q factors. After establishing
non-volatile tuning in individual LTOI devices, we next use the retained LT phase state as a functional
resource for active photonic circuits. At the device level, non-volatile phase reconfiguration provides
zero-static-power operating-point control of a \(>110~\mathrm{GHz}\) electro-optic modulator and trims
a cascaded modulator to a \(59.3~\mathrm{dB}\) extinction ratio. At the system level, we use image edge
detection as an example to show the potential of the LTOI platform for edge-intelligence processors.
The resulting image-edge-detection chip achieves a measured system-level energy efficiency of
\(3.48~\mathrm{TOPS\,W^{-1}}\) with zero on-chip static tuning power. Together, these demonstrations connect the
material mechanism, device-level reconfigurability and system-level operation of non-volatile LTOI
photonics.

\section{Results}
\label{results}

\subsection{Non-volatile refractive-index programming in LTOI}
\label{non-volatile-refractive-index-programming-in-ltoi}

We first establish the physical mechanism of non-volatile refractive-index programming in congruent
LTOI.
As shown in Fig. 1a, commercial lithium tantalate is typically grown with a congruent rather than a
strictly stoichiometric composition.\cite{Wang2024Nature,Kim2001DomainReversal} The resulting lithium deficiency is compensated primarily by
lithium vacancies and tantalum antisite defects, which are proposed to form defect-dipole complexes.
Within this model, the defect dipoles align with the spontaneous polarization of the host ferroelectric
domains at equilibrium and generate a built-in electric field
(Fig. 1a(ii)).\cite{Gopalan1996InternalField,Kim2001DomainReversal} When an external electric field switches ferroelectric domains
at room temperature, the local electro-optic coefficient changes sign within the switched fraction.
The effective electro-optic coefficient of the programmed region therefore changes. In the proposed
model, defect-dipole reorientation requires lithium vacancies to rearrange around the
\(\mathrm{Ta}_{\mathrm{Li}}\) antisites. Their limited room-temperature mobility prevents complete
rearrangement during domain reversal.\cite{Kim2001DomainReversal} Most defect dipoles therefore do not
fully follow the reversed ferroelectric domains. The defect-dipole-induced built-in field consequently
retains its initial direction and most of its magnitude in the programmed region (Fig. 1a(iii)).
This retained field interacts with the modified electro-optic
coefficient to produce a refractive-index change through the Pockels effect. After the external field
is removed, the retained switched-domain configuration preserves the modified electro-optic
coefficient, making the corresponding Pockels-induced refractive-index change non-volatile.
High-temperature annealing increases lithium-vacancy mobility and allows the vacancy configurations around
\(\mathrm{Ta}_{\mathrm{Li}}\) antisites to relax towards equilibrium with the retained domain
polarization. This process restores the equilibrium relative orientation and erases the programmed
refractive-index change (Fig. 1a(iv)).\cite{Kim2001DomainReversal} The refractive index therefore varies gradually with
the fraction of switched domains (Fig. 1d). The asymmetry between the positive and negative coercive
fields in the hysteresis loop of congruent LT provides further evidence for the defect-dipole-induced
built-in field.\cite{Gopalan1996InternalField,Kim2001DomainReversal}

\begin{figure}[p]
\centering
\captionsetup{font=small,skip=4pt}
\includegraphics[width=0.78\linewidth]{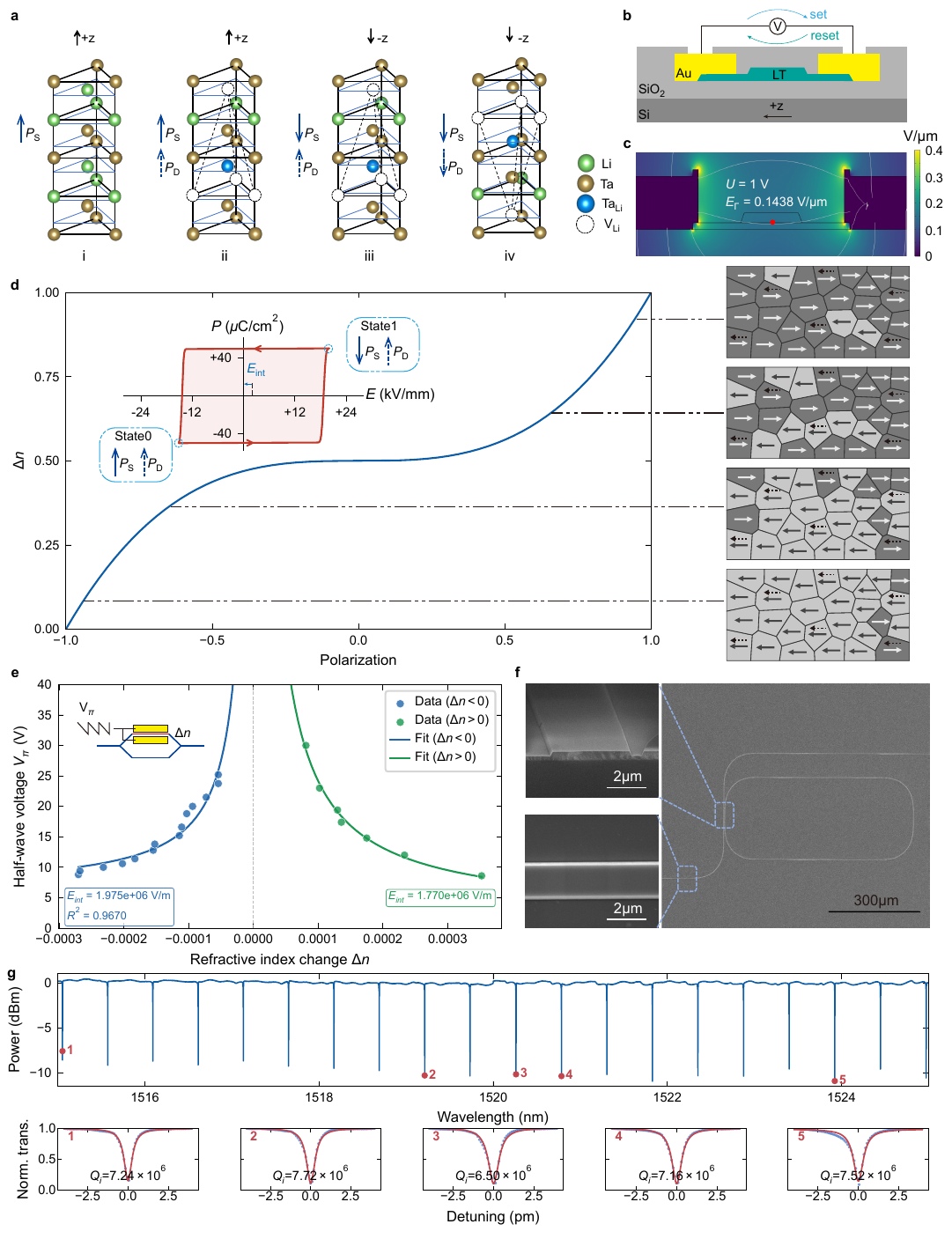}
\caption{\textbf{Defect-dipole-assisted non-volatile refractive-index programming in LTOI.}
\textbf{a}, LT unit-cell configurations. The \(+z\) axis is the polarization direction of the
as-received wafer; \(P_{\mathrm{S}}\) denotes domain polarization. (i) Stoichiometric LT. (ii) Congruent LT, where
\(\mathrm{Ta}_{\mathrm{Li}}\) antisites and \(V_{\mathrm{Li}}\) vacancies form defect dipoles with
\(P_{\mathrm{D}}\parallel P_{\mathrm{S}}\).\cite{Kim2001DomainReversal} (iii) Poling reverses
\(P_{\mathrm{S}}\), whereas limited \(V_{\mathrm{Li}}\) mobility prevents complete
\(P_{\mathrm{D}}\) reorientation. The built-in field therefore largely retains its initial direction
and magnitude. Domain reversal changes the effective electro-optic coefficient. The retained field
interacts with this modified coefficient to produce a non-volatile refractive-index change through
the Pockels effect. (iv) Annealing above
\(160\,^{\circ}\mathrm{C}\) for \(30~\mathrm{min}\) realigns
\(P_{\mathrm{D}}\) with \(P_{\mathrm{S}}\). \textbf{b}, Cross-section of the tuning region; slab
removal isolates domain switching in the ridge. \textbf{c}, Simulated electric-field distribution.
\textbf{d}, Normalized non-volatile \(\Delta n\) as the domain state evolves from initial
(\(-1\)) to fully reversed (\(1\)). The inset shows the hysteresis asymmetry and
\(E_{\mathrm{int}}\); solid and dashed arrows denote domain and defect-dipole orientations.
\textbf{e}, \(V_{\pi}\) versus \(\Delta n\). Fits yield
\(E_{\mathrm{int}}=1.97~\mathrm{kV\,mm^{-1}}\) for \(\Delta n<0\) and
\(1.77~\mathrm{kV\,mm^{-1}}\) for \(\Delta n>0\); inset, test MZI. \textbf{f}, Scanning electron
micrographs of the microring and waveguide. \textbf{g}, Microring spectrum and Lorentzian fits used
to extract intrinsic quality factors and propagation loss.}
\label{fig:fig1}
\end{figure}

The non-volatile refractive-index response can be described by an electro-optic model that incorporates
the built-in field generated by the defect dipoles. In the initial state, the defect dipoles are aligned
with the polarization direction of the ferroelectric domains. For the TE mode of an x-cut LTOI
waveguide, the extraordinary refractive index is the sum of its intrinsic value and the Pockels
contribution induced by this built-in field:

\[n_{e1} = n_{e}' - \frac{1}{2}{n_{e}'}^{3}\gamma_{33}E_{int}\]

Here, \(n_{e}'\) is the intrinsic extraordinary refractive index, \(\gamma_{33}\) is the electro-optic
coefficient and \(E_{int}\) is the effective built-in field generated by the defect dipoles. Applying 
an external electric field to the programmed region reverses a fraction \(a\) of the ferroelectric domains. 
These reversed domains contribute an effective Pockels response of opposite sign. The resulting 
extraordinary refractive index is:

\[n_{e2} = n_{e}' - \frac{1}{2}(1 - a){n_{e}'}^{3}\gamma_{33}E_{int} +
\frac{1}{2}a{n_{e}'}^{3}\gamma_{33}E_{int}\]

The refractive-index change is therefore

\[\Delta n_{e} = an_{e}^{3}\gamma_{33}E_{int}\]

The asymmetry between the positive and negative coercive fields reported for congruent LT
corresponds to an estimated built-in field of \(E_{int} \approx 2.1\) kV/mm.\cite{Gopalan1996InternalField,Kim2001DomainReversal}
Combined with the electro-optic response of LT, this estimate gives a bulk refractive-index change of \(0.613 \times
10^{-3}\). Accounting for optical confinement in the waveguide geometry, the corresponding
effective-index change is approximately \(0.55 \times 10^{-3}\) for the fundamental TE mode. To
stabilize the programmed domain configuration, we designed a dedicated poling region in LTOI. As
shown in Fig. 1b, the 400-nm-thick x-cut LTOI film is etched to a depth of 240 nm to form a rib
waveguide. The resulting waveguide supports the fundamental TE mode, whose optical field overlaps the
programmed region. The LT slab is locally removed on both sides of the active region. This geometry
reduces the influence of neighbouring unprogrammed domains. The electrodes are placed in direct contact 
with, or close to, the LT film to generate
a sufficiently strong poling field near the waveguide. Maxwell 2D electrostatic simulations (Fig. 1c)
show the cross-sectional electric-field distribution in the programmed region and quantify the
dependence of the waveguide field strength \(E_{\Gamma}\) on the applied voltage \(V\).

To test the defect-dipole model in LTOI waveguides, we measure the relation between the non-volatile
effective-index change and the half-wave voltage in an asymmetric Mach-Zehnder interferometer.
Partial domain switching in one arm changes both its static optical phase and its effective
electro-optic coefficient. We extract the static phase shift from the MZI spectral shift after removing
the programming field. To determine the effective electro-optic coefficient, we apply a 1-kHz
triangular voltage waveform to the non-volatile electrode and record the sinusoidal MZI response at
1550 nm. This measurement yields the half-wave voltage \(V_{\pi}\) of the programmed arm. Together,
these measurements establish the relation between \(\Delta n_{TE}\) and \(V_{\pi}\) for the programmed
arm.

\[\Delta n_{TE} = \frac{\pm \lambda}{1 +
\frac{{n_{e}'}^{2}\gamma_{33}\Gamma}{{n_{e}'}^{2}\gamma_{33}E_{int} + k}V_{\pi}}\]

Here, \(\lambda\) is the optical wavelength in the MZI, and \(\Gamma\) is the electric-field conversion
factor extracted from electrostatic simulations. It quantifies the waveguide electric field per unit
applied voltage. Because the electric field varies only weakly across the optical mode, \(\Gamma\) is
treated as a constant. The fitting parameter \(k\) accounts for a possible strain-induced
refractive-index contribution from domain reversal. We measured \(\Delta n_{TE}\) and \(V_{\pi}\) and
fitted the data to the equation above (Fig. 1e). The fit yields \(E_{int} \approx 1.97\) kV/mm when the
original domain orientation dominates and \(E_{int} \approx 1.77\) kV/mm when the reversed orientation
dominates. The fitted value of \(k\) is approximately \(5 \times 10^{-5}\). The lower fitted
\(E_{int}\) in the reversed orientation is consistent with partial field-driven reorientation of the
defect complexes during poling. The limited mobility of lithium vacancies prevents complete
rearrangement, so the built-in field retains its initial direction and most of its magnitude.
Cumulative depolarization during domain
switching may also contribute to this difference. The small value of \(k\) indicates that strain
contributes little to the non-volatile refractive-index change. Instead, the response is dominated by
the interaction between the retained built-in field and the domain-dependent effective
electro-optic coefficient.

Low propagation loss is essential for large-scale PICs.\cite{Bogaerts2020Programmable} We therefore fabricated ultralow-loss
waveguides in thin-film lithium tantalate (Fig. 1f). Electron-beam lithography, optimized dry etching
and KOH-based wet cleaning yielded LTOI rib waveguides with smooth sidewalls. We extracted the
propagation loss from the intrinsic quality factors of LTOI microring resonators (Fig. 1g). The
measured transmission spectra yield an intrinsic quality factor above \(7 \times 10^{6}\),
corresponding to a propagation loss of approximately \(0.05\)--\(0.06~\mathrm{dB\,cm^{-1}}\). Such low
loss is advantageous for applications such as optical delay lines, large-scale optical switch
matrices and Kerr-frequency-comb generation.\cite{Wang2024Nature} The coexistence of low-loss waveguiding and
defect-dipole-assisted non-volatile tuning motivates a quantitative assessment of tuning efficiency,
phase precision, retention and endurance.

\subsection{LTOI non-volatile tuning: performance and stability}
\label{ltoi-non-volatile-tuning-performance-and-stability}

Having established the non-volatile mechanism and its compatibility with low-loss waveguides, we next
quantify the tuning performance in asymmetric MZIs and microring resonators. LTOI non-volatile tuning
provides multilevel refractive-index control together with long-term retention, cycling endurance and
low optical loss. The
asymmetric MZI (Fig. 2a,b) contains a rib waveguide with a top width of \(2~\mu\mathrm{m}\) in the tuning
region. Local slab etching leaves a \(7~\mu\mathrm{m}\)-wide LT region, and the electrode gap is
\(6~\mu\mathrm{m}\). The electrodes directly contact the retained LT region. Trapezoidal voltage pulses
above 100 V switch the ferroelectric domains in either arm and increase the local waveguide
refractive index. Programming the long and short arms shifts the MZI interference spectrum to longer
and shorter wavelengths, respectively. When the long arm is programmed, optimized voltage pulses
applied to its 1-mm-long non-volatile tuning region produce non-volatile spectral
redshifts (Fig. 2c). Domain switching begins at 100 V and reaches full reversal at 127.5 V, yielding a
non-volatile phase shift of \(1.1\pi\).

\begin{figure}[p]
\centering
\captionsetup{font=small,skip=4pt}
\includegraphics[width=0.78\linewidth]{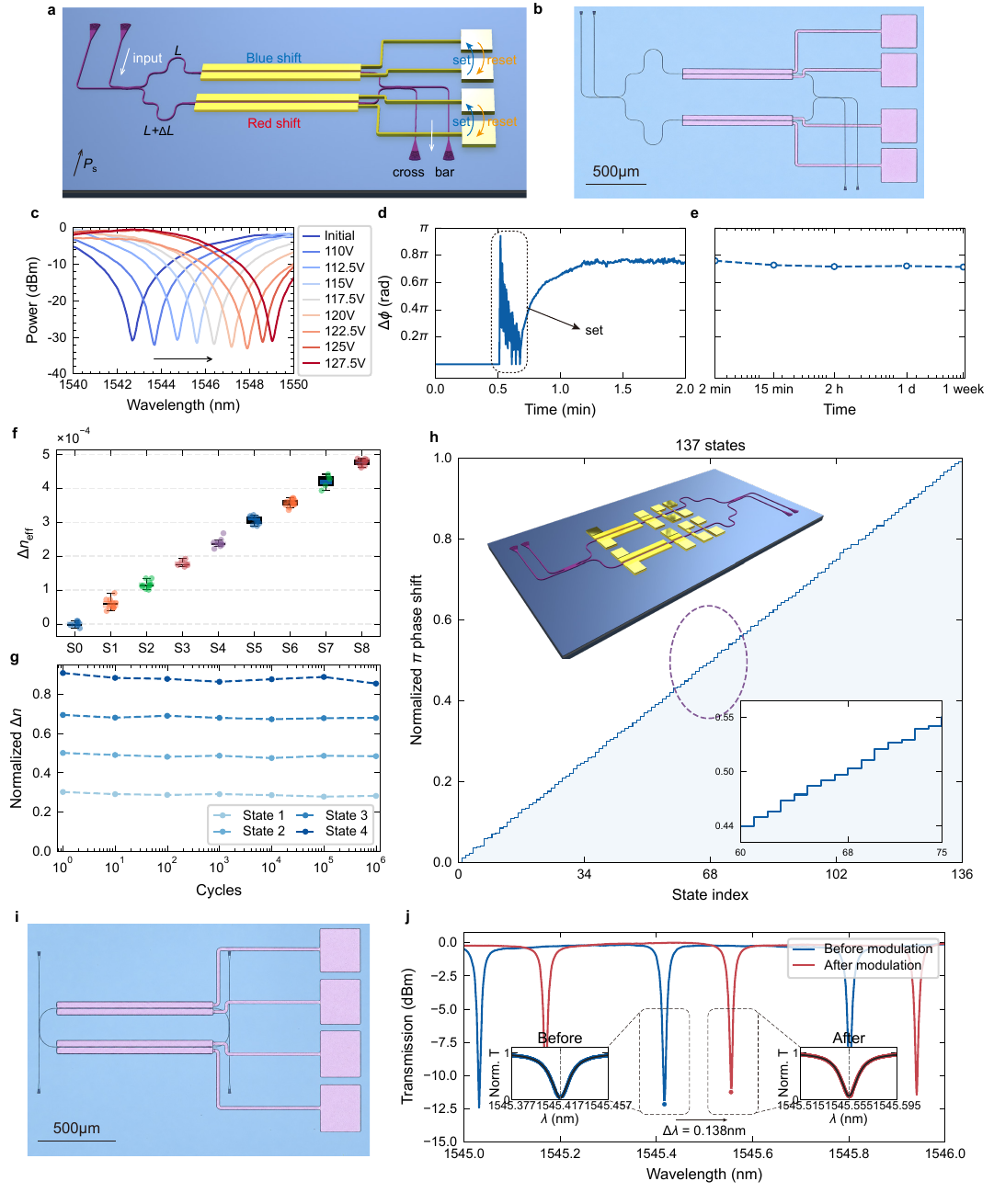}
\caption{\textbf{Multilevel, stable and low-loss non-volatile tuning in LTOI devices.}
\textbf{a}, Asymmetric MZI for characterizing non-volatile index changes; independent electrodes tune
the two arms in opposite directions. \textbf{b}, Optical micrograph. \textbf{c}, Transmission spectra
of programmed states. \textbf{d}, Phase evolution over \(2~\mathrm{min}\), including the write
transient. \textbf{e}, Retention from \(2~\mathrm{min}\) to one week. \textbf{f}, Effective-index
changes from \(S_0\) to \(S_1\)--\(S_8\). \textbf{g}, Four programmed states over
\(10^{6}\) forward--reverse poling cycles. \textbf{h}, Phase staircase from a four-section weighted
electrode. The device resolves 137 positions across \(\pi\), with an average step of
\(\sim0.007\pi\); inset, enlarged steps. \textbf{i}, Microring with non-volatile tuning electrodes.
\textbf{j}, Microring spectra before and after tuning, showing a resonance redshift; Lorentzian fits
show that the quality factor remains essentially unchanged.}
\label{fig:fig2}
\end{figure}

To resolve the temporal response of ferroelectric switching in LT, we monitor the MZI output power
with a high-speed optical power meter, beginning before the programming voltage is applied. From
these measurements, we extract the phase evolution over 2 min and one week. The first-minute phase
trace of the modulated arm (Fig. 2d) shows pronounced fluctuations. We attribute these fluctuations
to the charging and discharging of the electrode capacitance and to depolarization during domain
reversal. To assess long-term stability, we track the phase of the modulated MZI arm for one week on
a logarithmic time axis (Fig. 2e). Across the sampled time points, the measured phase variation remains
within 5\% of the programmed phase shift. The programmed state requires
no holding voltage and therefore consumes zero static electrical power. This behaviour contrasts
with volatile thermo-optic tuning and continuously biased control, and can reduce the power
consumption of large arrays of optical phase shifters.\cite{PerezLopez2020SelfConfiguration,Padmaraju2012Thermal}

The MZI supports multilevel non-volatile tuning through programmable refractive-index changes. With
the single-arm electrode configuration, the MZI exhibits eight clearly distinguishable programmed
states relative to the initial state across repeated experiments, corresponding to a 3-bit-equivalent
multilevel response (Fig. 2f). This result
shows that non-volatile programming is not limited to two optical states. The analogue-like response
arises from partial ferroelectric domain switching. By varying the pulse amplitude, width, polarity,
number, interval or waveform, the switched-domain fraction in the tuning region can be repeatedly
controlled.\cite{Kim2001DomainReversal}

We evaluate cycling endurance by reading four programmed states after different numbers of
forward--reverse poling cycles. Both the programmed tuning range and state distinguishability remain
statistically stable through \(10^{6}\) cycles (Fig. 2g).

Spatial weighting of the non-volatile phase-shifter segments increases the resolution of analogue
phase control. We use a four-section weighted electrode with a total length of
\(L = 2~\mathrm{mm}\). The section lengths are \(\frac{64}{85}L\), \(\frac{16}{85}L\),
\(\frac{4}{85}L\), and \(\frac{1}{85}L\). The longer sections provide coarse phase shifts, whereas
the shorter sections supply finer corrections. This response should be interpreted as analogue phase
trimming rather than digitally addressable phase states. Residual phase errors in the longer sections
alter the correction required from the shorter sections. Repeated tuning to the same target phase may
therefore produce different degrees of ferroelectric domain switching across the four sections, even
though the total optical phase shift reaches the same target. Experimentally, the phase staircase in
Fig. 2h resolves 137 phase positions across a \(\pi\) range, corresponding to an average phase-setting
step of \(\sim0.007\pi\). The residual non-uniformity of the staircase arises from small setting errors
in the non-volatile tuning state of individual electrode sections.

Non-volatile tuning remains compatible with low-loss resonant photonic devices. The microring
resonator in Fig. 2i incorporates a 1-mm-long non-volatile tuning region. Non-volatile tuning of this
region redshifts the resonance peak. The microring quality factor remains essentially unchanged
before and after tuning (Fig. 2j). This response does
not rely on an absorptive material phase or carrier redistribution. Instead, ferroelectric domain
switching changes the effective electro-optic coefficient. Because most defect dipoles retain their
initial orientation, the built-in field also retains its original direction and most of its
magnitude. The retained field interacts with the modified coefficient to sustain the non-volatile
refractive-index change. The largely preserved quality
factor therefore indicates negligible additional optical loss from non-volatile tuning.

Table 1 benchmarks LTOI against representative non-volatile photonic platforms across key performance
metrics. To our knowledge, this is the first monolithic LTOI platform without an added state-retentive
material to experimentally combine low propagation loss (\(\sim0.05~\mathrm{dB\,cm^{-1}}\)),
electro-optic modulation beyond \(110~\mathrm{GHz}\) and non-volatile phase tuning. With these
platform-level metrics
established, we next use non-volatile tuning to configure high-speed modulators and then extend the
same principle to system-level photonic processing.

\begin{table}[p]
\centering
\rotatebox[origin=c]{-90}{%
\begin{minipage}{0.94\textheight}
\centering
\fontsize{7}{8}\selectfont
\textbf{Table 1 \textbar{} Example comparison of non-volatile and programmable integrated photonic
platforms.}\par\medskip

\begin{tabular}{@{}
  >{\raggedright\arraybackslash}p{(\linewidth - 20\tabcolsep) * \real{0.0650}}
  >{\raggedright\arraybackslash}p{(\linewidth - 20\tabcolsep) * \real{0.0960}}
  >{\centering\arraybackslash}p{(\linewidth - 20\tabcolsep) * \real{0.0734}}
  >{\centering\arraybackslash}p{(\linewidth - 20\tabcolsep) * \real{0.1020}}
  >{\centering\arraybackslash}p{(\linewidth - 20\tabcolsep) * \real{0.0940}}
  >{\centering\arraybackslash}p{(\linewidth - 20\tabcolsep) * \real{0.1010}}
  >{\raggedright\arraybackslash}p{(\linewidth - 20\tabcolsep) * \real{0.0998}}
  >{\centering\arraybackslash}p{(\linewidth - 20\tabcolsep) * \real{0.0814}}
  >{\centering\arraybackslash}p{(\linewidth - 20\tabcolsep) * \real{0.0895}}
  >{\centering\arraybackslash}p{(\linewidth - 20\tabcolsep) * \real{0.0849}}
  >{\centering\arraybackslash}p{(\linewidth - 20\tabcolsep) * \real{0.1130}}@{}}
\toprule\noalign{}
\begin{minipage}[b]{\linewidth}\centering
\textbf{Platform}
\end{minipage} & \begin{minipage}[b]{\linewidth}\centering
\textbf{Non-volatile\newline mechanism}
\end{minipage} & \begin{minipage}[b]{\linewidth}\centering
\textbf{Non-volatile\newline refractive-\newline index change}
\end{minipage} & \begin{minipage}[b]{\linewidth}\centering
\textbf{Electro-optic\newline modulation speed}
\end{minipage} & \begin{minipage}[b]{\linewidth}\centering
\textbf{Programming\newline loss}
\end{minipage} & \begin{minipage}[b]{\linewidth}\centering
\textbf{Platform\newline optical loss}
\end{minipage} & \begin{minipage}[b]{\linewidth}\centering
\textbf{Fabrication\newline process}
\end{minipage} & \begin{minipage}[b]{\linewidth}\centering
\textbf{Non-volatile\newline tuning time}
\end{minipage} & \begin{minipage}[b]{\linewidth}\centering
\textbf{Endurance}
\end{minipage} & \begin{minipage}[b]{\linewidth}\centering
\textbf{Non-volatile\newline tuning energy}
\end{minipage} & \begin{minipage}[b]{\linewidth}\centering
\textbf{Demonstrated states\newline or phase positions}
\end{minipage} \\
\midrule\noalign{}
PCM/\newline Sb2S3\cite{Chen2023Sb2S3FiveBit} & Thermal\newline amorphous--\newline crystalline\newline switching & \(\Delta n = 1.8 \times 10^{-2}\) & N/A &
Single-MZI\newline insertion loss:\newline 1 dB & 240 dB/cm & Sb2S3-on-SOI\newline heterogeneous\newline integration & \(\mu\mathrm{s}\) scale
& 1,600 cycles & Set: 11.6 nJ (MZI); Reset: 197 nJ (MZI) & 32 \\
\hdashline
PCM/\newline Sb2Se3\cite{Yang2023NonVolatilePCM} & Thermal\newline amorphous--\newline crystalline\newline switching & \(\Delta n = 0.77\) &
N/A & 0.36 dB per\newline phase-shifter arm & 200 dB/cm &
Al2O3--Sb2Se3--\newline Al2O3 trilayer\newline on SOI & \(\mu\mathrm{s}\) scale & 10,000 cycles & Set: 105 nJ
(MZI); Reset: 1.2 \(\mu\mathrm{J}\) (MZI) & >32 (single);\newline >64 (array) \\
\hdashline
BTO\cite{GelerKremer2022BTO} & Ferroelectric\newline domain switching & \(\Delta n = 2 \times 10^{-3}\) & N/A & Negligible &
4.8 \(\pm\) 0.2 dB/cm & BTO-on-SOI\newline heterogeneous\newline integration & ns--\(\mu\mathrm{s}\) scale & \(10^{6}\) cycles & 4.6-26.7 pJ & 8 \\
\hdashline
BTO\cite{CatalaLahoz2026BTOFPPGA} & Ferroelectric\newline domain switching & N/A & High-speed electro-optic\newline switching speed\newline \(\sim\)80 ns & Negligible & 1.48 dB per device
& BTO-on-SOI\newline heterogeneous\newline integration & ns scale & N/A & 44.8 fJ & 16 \\
\hdashline
PZT\cite{Li2025PZTMemristor} & Ferroelectric\newline domain switching & \(\Delta n = 4.6 \times 10^{-3}\) & 48 Gbit/s (NRZ) &
Negligible & \(\sim\)1.9
dB/cm & Sol--gel PZT coating & 1-ms scale & >100,000 cycles &
12.3-126 pJ & 40 \\
\hdashline
PZT\cite{Li2026VersatilePZT} & Ferroelectric\newline domain switching & \(\Delta n = 5.3 \times 10^{-3}\) & 40 Gbit/s (NRZ) & Negligible & N/A & N/A & ms
scale & N/A & N/A & 40 \\
\hdashline
PZT\cite{Shu2026PZTMatrix} & Ferroelectric\newline domain switching & N/A & 40 Gbit/s (NRZ) & Negligible & 0.75-4 dB per device & Wafer-scale sol--gel\newline PZT coating & \(\mu\mathrm{s}\)--ms scale & >100,000 cycles\cite{Li2025PZTMemristor} & 120 pJ & 40 \\
\hdashline
MEMS\cite{Hu2025SWXMEMS} & Electrostatic\newline comb-drive\newline actuation & N/A &
N/A & N/A & 0.12-0.4 dB\newline per device (OFF);\newline 0.54-0.76 dB\newline per device (ON) & Monolithic SOI\newline integration & 1.2--3.5 \(\mu\mathrm{s}\) &
>\(10^{9}\) cycles & Sub-pJ & 2 \\
\hdashline
MEMS\cite{Hu2025NonvolatileMEMS} & van der Waals-\newline force latching
& N/A & N/A & N/A & 0.14-0.34 dB\newline per device (OFF);\newline 0.45-0.75 dB\newline per device (ON) & Monolithic SOI\newline integration & 2.2--7.4
\(\mu\mathrm{s}\) & >\(10^{6}\) cycles & \(\sim\)1 pJ & 2 \\
\hdashline
This work:\newline LTOI & Ferroelectric\newline domain switching\newline under a\newline defect-dipole field &
\(\Delta n = 0.613 \times 10^{-3}\) & >110 GHz & Negligible & \(\sim\)0.05 dB/cm & Monolithic LTOI
\newline integration & 100-ms scale & Through \(10^{6}\) cycles & Set: 2.6 \(\mu\mathrm{J}\) (MZI);\newline Reset: 3.06 \(\mu\mathrm{J}\) (MZI) & 8 programmed states\newline (excluding initial);\newline
137 phase positions\newline (weighted electrode) \\
\bottomrule
\end{tabular}
\end{minipage}%
}
\end{table}

\subsection{Zero-static-power bias control in a >110 GHz electro-optic modulator}
\label{zero-static-power-bias-control-in-a-110-ghz-electro-optic-modulator}

The optical power transfer function of a Mach--Zehnder modulator (MZM) is intrinsically nonlinear.
For an ideal balanced interferometer, it can be expressed as
\[
P_{\mathrm{out}}=\frac{P_{\mathrm{in}}}{2}\left[1+\cos\left(\pi
\frac{V_{\mathrm{RF}}}{V_{\pi}}+\phi_{\mathrm{bias}}+\phi_{0}\right)\right],
\]
where \(V_{\mathrm{RF}}\) is the radiofrequency (RF) modulation signal, \(V_{\pi}\) is the half-wave
voltage, \(\phi_{\mathrm{bias}}\) is the differential phase shift introduced by bias control and
\(\phi_{0}\) is the initial phase imbalance between the interferometer arms. The bias point determines the
linearity and harmonic response of the MZM. At the quadrature (Q) bias point, the MZM provides an
approximately linear response for high-speed optical communications and analogue photonic links. At
the peak (P) bias point,
constructive interference maximizes optical transmission and suppresses odd-order optical sidebands,
providing an even-order modulation response for microwave-photonic mixing and waveform synthesis. At
the null (N) bias point, destructive interference suppresses the optical carrier and provides a
low-transmission state for carrier-suppressed modulation, high-extinction optical pulse generation and
photonic microwave generation. Beyond these canonical operating points, arbitrary bias states enable
programmable control of carrier--sideband relations and, in suitable modulator architectures, support
single-sideband modulation, microwave-photonic signal processing and analogue photonic weighting.\cite{Shi2019BiasController,Marpaung2019IntegratedMWP}

Conventional electro-optic and thermo-optic bias stabilization generally requires a monitor
photodiode, feedback electronics and a continuously applied bias voltage or heating power.\cite{Shi2019BiasController,Padmaraju2012Thermal} Reliable
locking requires careful coordination of the optical tap ratio, photodetector dynamic range,
control-circuit resolution and feedback algorithm. The associated monitoring and control hardware
increases the system footprint, cost and power consumption. Large variations in input optical power
can push the monitor photodiode outside its operating range or distort the feedback error signal,
causing loss of bias lock, while
pilot-tone schemes may introduce low-frequency perturbations into the link.\cite{Shi2019BiasController,Ackerman2000PilotTone}

By contrast, non-volatile bias control consumes energy only during programming and retains the phase
state with zero holding power. Monitoring and feedback are therefore required only for programming or
recalibration, and variations in input optical power do not cause feedback-induced loss of the stored
state. This
set-and-retain operation reduces the cumulative power and control overhead of large-scale modulator
arrays and preserves programmed states across power interruptions.

To implement non-volatile bias control without compromising the high-speed modulation path, we
separate bias programming from RF modulation in an LTOI MZM comprising two \(2\times2\)
multimode-interference couplers, a \(6\)-mm-long travelling-wave modulation electrode and segmented
non-volatile electrodes on both interferometer arms (Fig. 3a). The travelling-wave electrode applies
the high-speed RF drive, whereas the non-volatile electrodes are activated only during bias
programming. With a total interaction length of \(2~\mathrm{mm}\), the non-volatile electrodes are
geometrically weighted in a length ratio of 1:4:16:64, enabling multiscale programming of the
differential phase. After the programming voltage is removed, the remanent ferroelectric-domain
configuration retains the programmed phase offset without a holding voltage.

\begin{figure}[p]
\centering
\captionsetup{font=small,skip=4pt}
\includegraphics[width=0.78\linewidth]{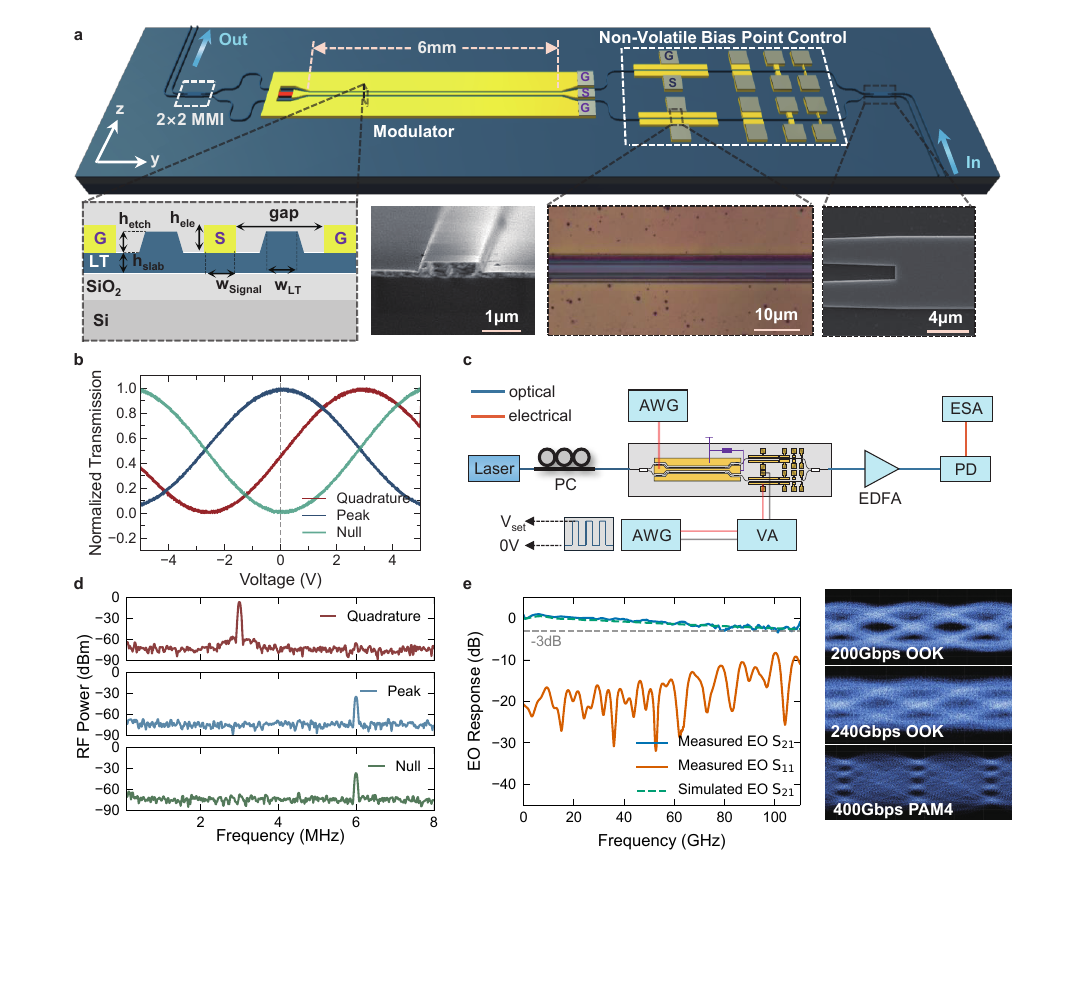}
\caption{\textbf{Zero-static-power bias control of an ultrahigh-speed LTOI Mach--Zehnder modulator.}
\textbf{a}, Asymmetric MZM combining a \(6\)-mm travelling-wave electrode with segmented
non-volatile bias electrodes; insets, electrode cross-section and device micrographs. \textbf{b},
Transmission at the programmed P, Q and N bias points. \textbf{c}, Set-up for bias programming and
modulation measurements. PC, polarization controller; AWG, arbitrary waveform generator; VA, voltage
amplifier; EDFA, erbium-doped fibre amplifier; PD, photodetector; ESA, electrical spectrum analyser.
\textbf{d}, RF spectra under a \(3~\mathrm{MHz}\) drive at the three bias points. \textbf{e}, Measured
and simulated electro-optic responses, with eye diagrams at \(200\) and
\(240~\mathrm{Gbit\,s^{-1}}\) OOK and \(400~\mathrm{Gbit\,s^{-1}}\) PAM-4.}
\label{fig:fig3}
\end{figure}

Non-volatile programming sets the MZM to the P, Q and N bias points at
\(1550~\mathrm{nm}\) (Fig. 3b). We verify the programmed bias states from their harmonic responses to a
\(3~\mathrm{MHz}\) sinusoidal drive (Fig. 3c). At the Q bias point, the second-harmonic component is more
than \(50~\mathrm{dB}\) below the fundamental, indicating precise non-volatile programming of this
point. At the P and N bias points, the photodetected component at the drive frequency is strongly
suppressed (Fig. 3d). We further assess the temporal stability of the Q bias point, for which the
second-harmonic suppression remains close to \(50~\mathrm{dB}\) over \(80~\mathrm{h}\) without an
applied holding voltage, confirming stable retention of the programmed bias state (Supplementary
Information).

Finally, we establish that non-volatile bias control is compatible with ultrahigh-speed modulation.
At the Q bias point, the \(6\)-mm-long travelling-wave MZM exhibits an electro-optic 3-dB bandwidth around
\(110~\mathrm{GHz}\) (Fig. 3e). Eye-diagram measurements further demonstrate
\(200\)- and \(240~\mathrm{Gbit\,s^{-1}}\) OOK operation and \(400~\mathrm{Gbit\,s^{-1}}\) PAM-4
operation (Fig. 3e). The experimental set-up and measurement procedures are described in the
Supplementary Information. Together, these measurements demonstrate the compatibility of non-volatile
bias programming with broadband and high-speed data modulation.

These results demonstrate precise non-volatile phase control of the LTOI modulator bias point, with no
static holding power and data rates up to \(400~\mathrm{Gbit\,s^{-1}}\). We next apply the same
phase-control capability to compensate interferometric imbalance and enhance the extinction ratio.

\subsection{Non-volatile extinction-ratio enhancement up to 60 dB}
\label{non-volatile-extinction-ratio-enhancement-up-to-60-db}

Ultrahigh-extinction-ratio electro-optic modulators are essential when residual off-state light raises
the noise floor or compromises measurement and control fidelity. In coherent distributed fibre sensing, leakage between probe
pulses produces a Rayleigh-scattered background that degrades the sensing range and spatial resolution;\cite{Ren2016FiniteER}
in quantum photonics, it can mask weak photon correlations and compromise the control of photon-number
statistics. An extinction ratio above \(50~\mathrm{dB}\) suppresses the residual optical power below one
part in \(10^{5}\), providing a practically important level for weak-signal detection and quantum-light
control. Experimentally, improving the pulse extinction ratio from \(20\) to \(50~\mathrm{dB}\) enhanced
spatial-crosstalk suppression by \(24.9~\mathrm{dB}\) in a distributed acoustic sensing system,\cite{Shen2024UltraHighERDAS} while a
\(51~\mathrm{dB}\) TFLN modulator enabled voltage-programmable control of photon-number statistics, with
the zero-delay second-order intensity correlation \(g^{(2)}(0)\) tuned from approximately 1.0 to 1.7.\cite{Bankwitz2026PhotonStatistics}

Achieving ultrahigh extinction requires tight matching of the optical amplitudes in the two
interferometer arms, because splitting-ratio errors and differential arm loss cannot be corrected by
null-bias control alone. An auxiliary MZI operating as a variable beam splitter can compensate for
these amplitude imbalances, but conventional implementations rely on continuously driven thermo-optic
or electro-optic tuning.\cite{Jin2019HighExtinction,Bankwitz2026PhotonStatistics} Here, the variable splitter is programmed using non-volatile electrodes,
allowing the balanced state to be retained after the writing voltage is removed. The preceding
measurements established multilevel non-volatile programming with a single electrode, while the shortest
segmented electrode provides a phase increment of \(0.0026\pi\). An extinction ratio above
\(50~\mathrm{dB}\) requires access to an approximately \(0.004\pi\)-wide phase window around the null
point, indicating that the measured phase granularity is sufficiently fine to support extinction ratios
above \(50~\mathrm{dB}\).

Accordingly, we integrate the non-volatile balancing interferometer with a high-speed travelling-wave
MZM (Fig. 4a). The first \(2\times2\) MMI splits the input power between the modulation arms, while two
additional MMIs and segmented non-volatile electrodes form the balancing interferometer. The
\(6\)-mm-long travelling-wave electrodes provide high-speed modulation. Simulations show that the MMI
exhibits a non-ideal optical power splitting ratio (Fig. 4b). For a representative amplitude-imbalance
state, tuning the phase of the programmable balancing interferometer redistributes the optical power
between the two paths and substantially enhances the extinction ratio of the high-speed MZM (Fig. 4c).
We experimentally program six states, S0--S5, and measure their transmission spectra
(Fig. 4d). At \(1553~\mathrm{nm}\), the extinction ratio increases from \(35.9\) to
\(59.3~\mathrm{dB}\), corresponding to an improvement of \(23.4~\mathrm{dB}\). The programmed amplitude
balance is retained without a DC holding voltage. Further theoretical details are provided in the
Supplementary Information.

\begin{figure}[p]
\centering
\captionsetup{font=small,skip=4pt}
\includegraphics[width=0.78\linewidth]{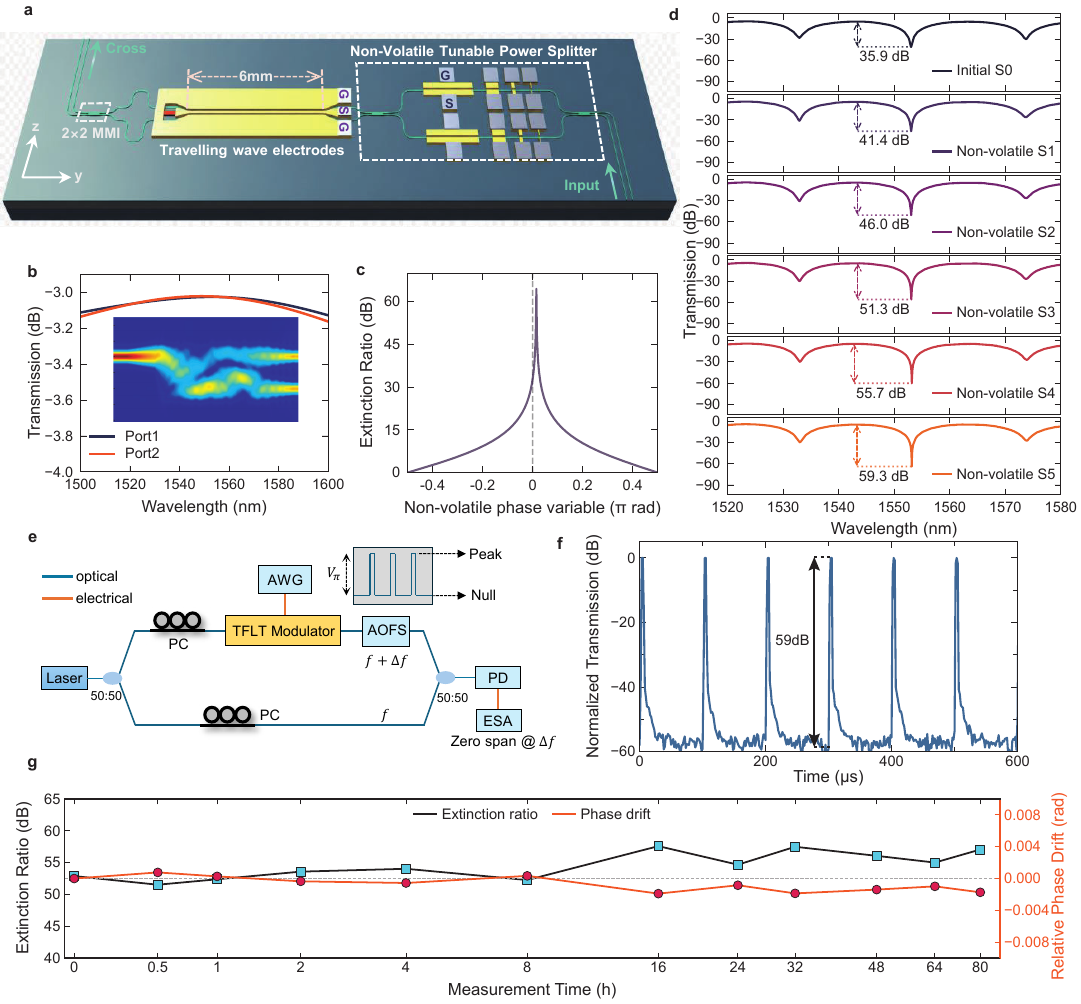}
\caption{\textbf{Non-volatile power balancing for an extinction ratio approaching 60 dB.}
\textbf{a}, Cascaded LTOI device combining a non-volatile tunable splitter with a \(6\)-mm
travelling-wave MZM. \textbf{b}, Simulated \(2\times2\) MMI output spectra and optical-field
distribution. \textbf{c}, Calculated extinction ratio versus programmed auxiliary-interferometer
phase. \textbf{d}, Spectra from the initial state \(S_0\) to \(S_1\)--\(S_5\), showing an extinction-ratio
increase from \(35.9\) to \(59.3~\mathrm{dB}\) at \(1553~\mathrm{nm}\). \textbf{e},
Self-heterodyne measurement set-up. PC, polarization controller; AWG, arbitrary waveform generator;
AOFS, acousto-optic frequency shifter; PD, photodetector; ESA, electrical spectrum analyser.
\textbf{f}, Temporal waveform with an extinction ratio of approximately \(59~\mathrm{dB}\).
\textbf{g}, Extinction ratio and equivalent phase drift over \(80~\mathrm{h}\) without reprogramming
or a DC bias.}
\label{fig:fig4}
\end{figure}

High-extinction optical pulses suppress inter-pulse leakage, thereby reducing background scattering in
distributed fibre sensing and unwanted photon populations in time-bin quantum systems.\cite{Ren2016FiniteER,Wang2022QKDExtinction} We therefore
examine whether the programmed high-extinction state is preserved during dynamic optical pulse
generation. The common-source optical heterodyne measurement set-up is shown in Fig. 4e. Continuous-wave
light at \(1553~\mathrm{nm}\) is split into signal and reference arms. The light in the signal arm is
modulated by a square-wave electrical signal with a 5\% duty cycle and a \(100~\mu\mathrm{s}\) period,
while an \(80~\mathrm{MHz}\) frequency shift is introduced for heterodyne detection before the two arms
are recombined. The resulting electrical beat note is recorded using an electrical spectrum analyser
operated in zero-span mode to recover the temporal waveform. The generated pulses exhibit an extinction
ratio of approximately \(59~\mathrm{dB}\),
consistent with the static value of \(59.3~\mathrm{dB}\) (Fig. 4f), confirming that the programmed
amplitude balance is preserved during dynamic modulation. A weak trailing-edge transient, approximately
\(40~\mathrm{dB}\) below the pulse maximum, may originate from the low-frequency electro-optic or
electrical response and could be reduced by pre-emphasis.\cite{Shen2024UltraHighERDAS}

Finally, we monitor the programmed high-extinction state for \(80~\mathrm{h}\) without further
non-volatile programming or an applied DC bias (Fig. 4g). Starting from \(52.88~\mathrm{dB}\), the
extinction ratio remains above \(51~\mathrm{dB}\). The observed variation likely arises from the strong
sensitivity of near-complete destructive interference to environmental and measurement fluctuations.
Even if the entire variation is attributed to phase drift, the corresponding deviation remains below
\(0.001\pi\), demonstrating stable retention of the programmed state.

These results validate precise and stable non-volatile phase control in high-speed LTOI modulators,
providing the basis for the system-level implementation described below.

\subsection{High-speed photonic chip for image edge detection enabled by non-volatile modulation}
\label{high-speed-photonic-chip-for-image-edge-detection-enabled-by-non-volatile-modulation}

Image edge detection is a key preprocessing step in computer vision. It extracts structural features
by identifying local changes in grey level between a candidate pixel and its neighbourhood.\cite{Canny1986EdgeDetection} Electronic
processors perform this task flexibly, but real-time high-resolution images require repeated data
movement and many pixel-wise operations. Photonic approaches can increase throughput by processing
optical signals in parallel or at high speed.\cite{Xu2021PhotonicAccelerator} However, many reported edge-detection chips rely on
thermally stabilized resonators, wavelength channels or analogue weights, which add static power,
control complexity and sensitivity to weight-setting errors.\cite{Xu2022PTFP,Bai2023Microcomb}

Here, we combine LTOI non-volatile tuning with high-speed electro-optic modulation in a system-level
photonic processor for image edge detection. In this chip, non-volatile tuning writes and
retains the operating point of each modulator, thereby storing the edge-detection criterion without
continuous tuning power. High-speed electro-optic modulation loads binary signals derived from
comparisons between neighbouring pixels. The cascaded modulator arrays therefore recast local edge
decisions as digital optical pattern matching, simplifying control while supporting high-speed and
energy-efficient computation.

Figure 5a shows how a single LTOI modulator functions as a configurable one-bit optical comparator.
Non-volatile phase tuning programs the modulator near the positive-slope linear point \(Q^{+}\), the
negative-slope linear point \(Q^{-}\), or the transmission extremum \(P\). At \(Q^{+}\), binary electrical
inputs of \(+1\) and \(-1\) produce high and low optical intensities, respectively. At \(Q^{-}\), the sign
of the electro-optic transfer-curve slope is reversed, thereby inverting this intensity mapping. Near
\(P\), the local slope approaches zero, making the output intensity largely insensitive to the input
sign. This condition implements the don't-care (\(X\)) state in the image-edge-detection criterion.
Thus, non-volatile programming of the operating point stores the comparison criterion, whereas the
high-speed electro-optic signal provides the input to be evaluated. Figure 5b shows that the same test
sequence produces distinct high-speed optical responses at \(Q^{+}\) and \(Q^{-}\). These
measurements confirm that a single modulator can implement the \(+1\), \(-1\), and \(X\) comparison
states.

\begin{figure}[p]
\centering
\captionsetup{font=small,skip=4pt}
\includegraphics[width=0.78\linewidth]{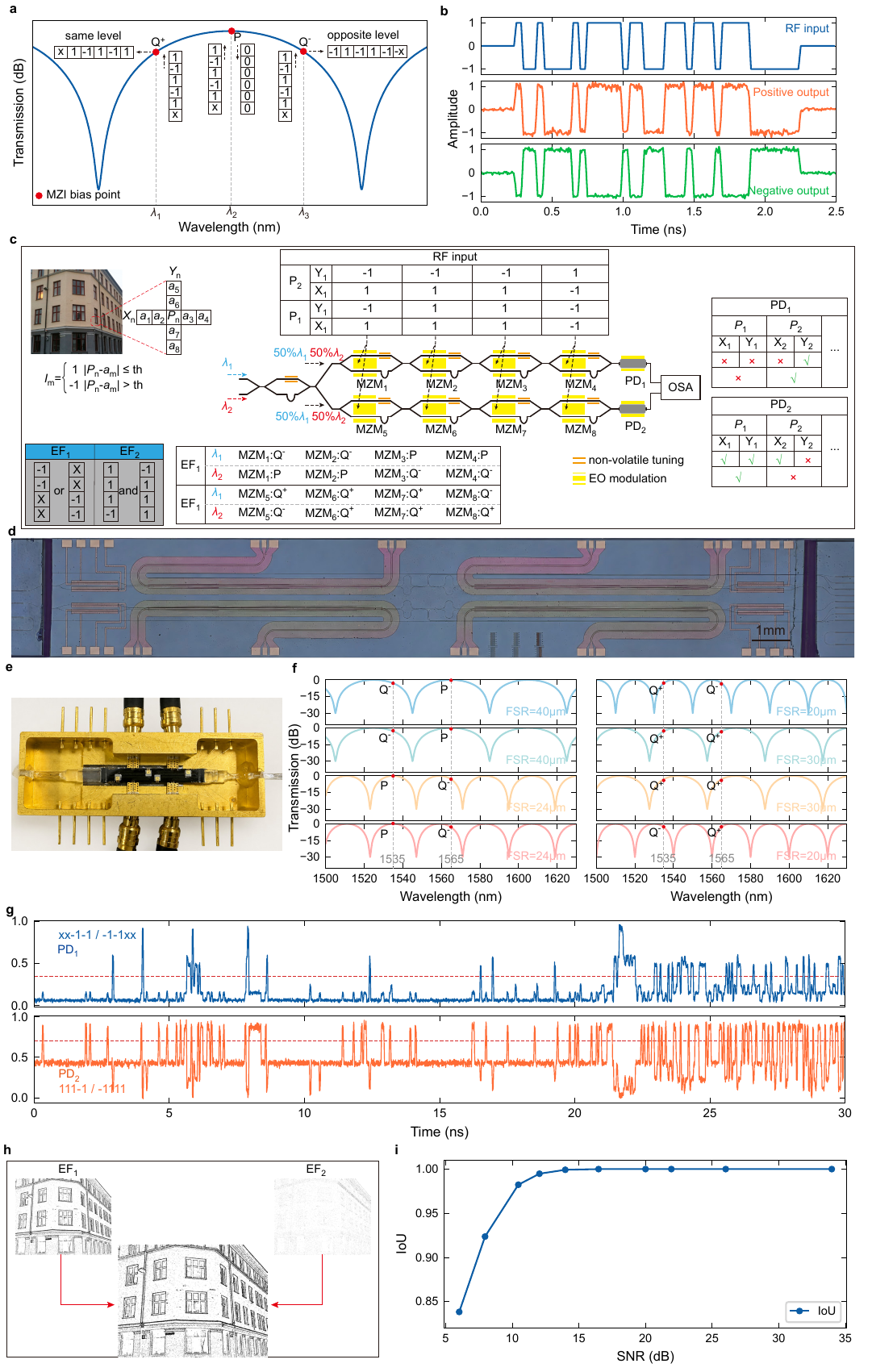}
\caption{\textbf{LTOI image-edge-detection chip combining non-volatile phase control with high-speed
electro-optic modulation.} \textbf{a}, Schematic transmission spectrum of an asymmetric MZI showing
the \(Q^{+}\), \(P\) and \(Q^{-}\) operating points and their mappings: \(Q^{+}\) preserves the sign,
\(Q^{-}\) inverts it and \(P\) represents the don't-care state \(X\). \textbf{b}, Input and output
traces at \(Q^{+}\) and \(Q^{-}\). \textbf{c}, Chip architecture. Non-volatile tuning stores the
\(+1\), \(-1\) and \(X\) criteria, while high-speed modulation encodes neighbouring pixels. Two
wavelengths and two \(1\times4\) modulator chains evaluate \(EF_1\) and \(EF_2\) in parallel.
\textbf{d}, Optical micrograph of a modulator chain. \textbf{e}, Packaged chip. \textbf{f}, Transfer
spectra of both chains after non-volatile alignment. \textbf{g}, \(30~\mathrm{ns}\) detector traces
for \(EF_1\) and \(EF_2\), with thresholds of 0.35 and 0.7. Simultaneous
\(XX{-}1{-}1\) and \({-}1{-}1XX\) matches give PD1 twice the amplitude range of PD2. \textbf{h}, Edge
map from the union of the two decisions. \textbf{i}, IoU under zero-mean Gaussian input noise; it
remains unity above an SNR of approximately \(15~\mathrm{dB}\).}
\label{fig:fig5}
\end{figure}

Building on this configurable comparator, we cascade LTOI modulators to perform local pattern matching
for image edge detection. We use a variant of the smallest univalue segment assimilating nucleus
(SUSAN) algorithm, in which the conventional circular mask is replaced by a cross-shaped mask.\cite{Chen2025FeFETEdge} Each
candidate pixel is compared with neighbouring pixels within this mask to determine grey-level
similarity. The comparison outcomes are encoded as binary arrays of \(+1\) and
\(-1\). The \(+1\), \(-1\), and \(X\) entries of the edge criteria EF1 and EF2 are programmed as the
\(Q^{+}\), \(Q^{-}\), and \(P\) operating points, respectively. This mapping recasts image edge
detection as an equality test. The test compares high-speed binary signals from the pixel neighbourhood
with optical criteria stored by non-volatile tuning. The cascaded modulator chain produces a
high-intensity optical response only when the input array matches the programmed criterion.

To increase hardware utilization, we incorporate wavelength-division multiplexing (WDM) into the
optical comparison architecture. As shown in Fig. 5c,f, two optical carriers at \(\lambda_1\) and
\(\lambda_2\) propagate simultaneously through each asymmetric-MZI comparator. Its wavelength-dependent
phase response places the two carriers at different operating points, allowing a single physical
modulator to encode a distinct comparison criterion at each wavelength.
Consequently, two \(1 \times 4\) cascaded-modulator chains can simultaneously evaluate multiple
\(1 \times 4\) criteria from EF1 and EF2, reducing the number of modulators needed for each edge
decision. At the input, a \(2 \times 2\) asymmetric MZI containing a non-volatile phase-tuning arm
routes the two carriers to the modulator chains. The free spectral range (FSR) of this router is
designed to equal twice the wavelength separation, \(2|\lambda_2-\lambda_1|\). Non-volatile phase
tuning of the arm shifts the transfer spectrum until both carriers align with 3-dB splitting points.
When \(\lambda_1\) and \(\lambda_2\) are launched into separate input ports, each carrier is therefore
divided equally between the two output ports feeding the serial modulator chains. Four high-speed RF
signals carrying the \(X_n\) and \(Y_n\) neighbourhood features in pixel order are then applied in
parallel to the travelling-wave electrodes. Finally, simple logic combines the high and low detector
outputs to produce the edge decision.

Owing to limitations of the available packaging and test setup, the experiment uses two separate
\(1 \times 4\) LTOI modulator chains (Fig. 5d). Off-chip wavelength-division multiplexers and optical
splitters combine and distribute the \(\lambda_1\) and \(\lambda_2\) carriers. The cascaded-modulator
chip is coupled to a fibre array through trident edge couplers. Each modulator is \(6~\mathrm{mm}\) long 
and has a half-wave voltage--length product (\(V_{\pi}L\)) of \(3.1~\mathrm{V}\!\cdot\!\mathrm{cm}\). 
Before packaging, the measured on-chip 3-dB electro-optic bandwidth is \(103~\mathrm{GHz}\). Figure 5e 
shows the modulator chip housed in a butterfly package. This co-integration provides a hardware platform 
for system-level high-speed optical computing.

To validate system-level operation, we evaluate the LTOI photonic chip using standard test images.
The available test setup limits the experimental input data rate to
\(20~\mathrm{Gbit\,s^{-1}}\). The reconstructed edge map agrees with the computer-simulated result
(Fig. 5h), yielding an edge-recognition accuracy of \(100\%\). This agreement shows that the criteria
encoded by the non-volatile operating points remain stable during high-speed electro-optic operation
and can support the complete image-processing task. The architecture uses binary pattern matching and
discrete high/low decisions instead of precise continuous analogue weights. This discrete architecture
is therefore less sensitive to device errors, phase drift, accumulated loss and readout noise. To
assess input-noise tolerance, we add zero-mean Gaussian noise of varying strength to the input image or
feature sequence and quantify the noise level using the signal-to-noise ratio (SNR). We evaluate
edge-reconstruction quality using the intersection over union (IoU) between the chip output and the
reference edge map (Fig. 5i). The IoU remains unity for SNRs above approximately
\(15~\mathrm{dB}\), indicating exact recovery of the tested edge map in this noise range. Even when the
SNR decreases to \(10.46~\mathrm{dB}\), the IoU remains \(0.982\), corresponding to a mismatch of less
than \(2\%\) relative to the union of the detected and reference edge pixels. This near-unity overlap
shows that the non-volatile binary-decision architecture preserves a large decision margin under input
noise.

We evaluate the system-level performance of the non-volatile LTOI chip in Table 2. At the measured
input rate of \(20~\mathrm{Gbit\,s^{-1}}\), the chip processes \(10~\mathrm{Gpixel\,s^{-1}}\) with a
computational throughput of \(0.32~\mathrm{TOPS}\) and a system-level computing efficiency of
\(3.48~\mathrm{TOPS\,W^{-1}}\). The eight LT non-volatile tuning electrodes set the optical criteria
and require no holding power after configuration. If they were replaced by thermo-optic phase
shifters with the \(P_{\pi}\approx55~\mathrm{mW}\) value measured for our earlier LTOI devices, driving
all eight at \(P_{\pi}\) would add \(0.44~\mathrm{W}\) of static power. At the same throughput, the
computing efficiency would fall to approximately \(0.60~\mathrm{TOPS\,W^{-1}}\). This comparison
quantifies the system-level benefit of non-volatile phase control. The power-accounting boundary for
the measured efficiency includes the optical output power of the two lasers, the
\(50~\Omega\)-terminated modulators, the data converters and the photodetector bias power, while
excluding laser wall-plug inefficiency. For the tested image, the deterministic optical
criteria reproduce the reference edge map exactly in the absence of added noise, yielding \(100\%\)
edge-map agreement. For an idealized operating rate of \(200~\mathrm{Gbit\,s^{-1}}\), the projected
processing speed reaches \(100~\mathrm{Gpixel\,s^{-1}}\), corresponding to \(3.2~\mathrm{TOPS}\).
Assuming four 2-bit DACs and two 2-bit ADCs whose power is scaled using reported data-converter figures
of merit, the corresponding projected computing efficiency is \(120.4~\mathrm{TOPS\,W^{-1}}\). This
converter-core projection excludes the power required for transimpedance amplification, clock and data
distribution, and laser wall-plug loss.
The present setup nevertheless uses two off-chip C-band lasers, an arbitrary waveform generator and
photodetectors, with simple decision logic executed by an FPGA or host computer. Integrating the light
sources, detectors, and drive and readout electronics could further reduce system-level power
consumption. Direct comparison across edge-detection hardware is affected by differences in algorithms,
datasets, system boundaries and power accounting; Table 2 therefore provides context rather than an
absolute ranking.

\begin{table}[p]
\centering
\rotatebox[origin=c]{-90}{%
\begin{minipage}{0.94\textheight}
\centering
\scriptsize
\textbf{Table 2 \textbar{} Comparison of representative image edge-detection chips.}\par\medskip

\begin{tabular}{@{}
  >{\centering\arraybackslash}p{(\linewidth - 8\tabcolsep) * \real{0.2000}}
  >{\centering\arraybackslash}p{(\linewidth - 8\tabcolsep) * \real{0.2000}}
  >{\centering\arraybackslash}p{(\linewidth - 8\tabcolsep) * \real{0.2000}}
  >{\centering\arraybackslash}p{(\linewidth - 8\tabcolsep) * \real{0.2000}}
  >{\raggedright\arraybackslash}p{(\linewidth - 8\tabcolsep) * \real{0.2000}}@{}}
\toprule\noalign{}
\multicolumn{1}{c}{\textbf{Platform}} & \textbf{Edge-detection speed\newline (pixels per second)} & \textbf{Edge-recognition\newline
accuracy\newline (no added noise)} & \textbf{Computing\newline efficiency} & \multicolumn{1}{c}{\begin{tabular}[c]{@{}c@{}}\textbf{Holding/tuning}\\\textbf{power}\end{tabular}} \\
\midrule\noalign{}
CMOS\cite{Jin2020EdgeCIS} & 16.6 Mpixel/s & 94.96\% & FoM = 56.7
pJ/pixel/frame & 9.4 mW \\
\hdashline
CMOS\cite{Lee2025LowPowerEdgeCIS} & 3.4 Mpixel/s & >90\% & FoM = 28.1 pJ/pixel/frame & 1.52 mW
\\
\hdashline
SOI\cite{Xu2022PTFP} & 20 Gpixel/s & Action-recognition\newline accuracy: 97.9\% & N/A &
On-chip WDM thermal-tuning\newline electrodes and microring\newline heater electrodes \\
\hdashline
heterogeneous integration\newline on SOI\cite{Bai2023Microcomb} & 17 Gpixel/s & N/A & 0.2 TOPS/W (experiment);\newline 2.38
TOPS/W (projected) & 238 mW from on-chip\newline thermal-tuning electrodes;\newline 89.5 W total system power \\
\hdashline
This work & Measured: 10 Gpixel/s;\newline projected: 100
Gpixel/s & 100\% & 3.48 TOPS/W ;\newline FoM = 9.20 pJ/pixel/frame (measured, 20 Gbit/s);\newline 120.4 TOPS/W ;
\newline FoM = 0.266 pJ/pixel/frame (projected, 200 Gbit/s)& 0 W on-chip\newline static tuning power \\
\bottomrule
\end{tabular}
\end{minipage}%
}
\end{table}

Together, these measurements validate the system-level role of non-volatile LTOI photonics.
Non-volatile tuning stores the optical comparison criteria, high-speed electro-optic modulation loads
real-time input data, and cascaded modulator arrays perform optical-domain pattern matching with
strong tolerance to input noise and without on-chip static tuning power. This demonstration completes
the progression of the Results from defect-dipole-mediated material non-volatility to low-loss phase
tuning, device-level reconfiguration and circuit-level operation. It shows that LTOI can retain optical
configuration while combining low-loss optical paths with high-speed electro-optic modulation in a
single integrated photonics platform.

\section{Discussion}
\label{discussion}

Together, these results establish LTOI as a non-volatile photonic platform that combines three
capabilities that are difficult to achieve in a single material system. First, LTOI supports low-loss
waveguides, with a propagation loss of \(\sim0.05~\mathrm{dB\,cm^{-1}}\), so programmed circuits can
scale without a large passive-loss penalty.\cite{Wang2024Nature} Second, its Pockels effect enables high-speed
electro-optic modulation with bandwidths beyond \(110~\mathrm{GHz}\), providing a direct path to
high-throughput signal processing.\cite{Wang2024Optica} Third, ferroelectric programming, interpreted
through the proposed defect-dipole model, stores optical phase states after the programming voltage is
removed, enabling reconfigurable photonic circuits without on-chip static tuning
power.\cite{Gopalan1996InternalField,Kim2001DomainReversal}

The benchmark in Table 1 shows that the trade-offs of non-volatile photonics are set by the
underlying storage mechanism. PCMs provide large refractive-index changes because the transition
between amorphous and crystalline phases strongly modifies atomic order, bonding, electronic
polarizability and permittivity.\cite{Wuttig2017PCM} The same transition, however, can also change optical absorption,
which remains a central limitation for many PCMs in low-loss photonic devices.\cite{Wuttig2017PCM} Repeated phase transitions can
further cause thermal and structural degradation, leading to optical-state drift and endurance
failure.\cite{Popescu2025PCMFailure} MEMS photonic devices avoid material absorption and can provide low-loss latching, but
reported devices generally offer discrete on--off tuning.\cite{Hu2025NonvolatileMEMS} This makes them less suited to dense and
high-precision waveguide phase control. Their movable structures also add device complexity and
require mechanical packaging and reliability engineering.\cite{Quack2023PhotonicMEMS} BTO offers very
low programming energies\cite{GelerKremer2022BTO} and provides some of the fastest non-volatile switching
among the ferroelectric platforms compared here.\cite{CatalaLahoz2026BTOFPPGA} In the reported devices compared here, however, BTO
and PZT still face material- and integration-related loss penalties.\cite{GelerKremer2022BTO,Li2025PZTMemristor} Their large dielectric
permittivity can also complicate microwave--optical velocity matching and limit the bandwidth of
Mach--Zehnder electro-optic modulators.\cite{Chelladurai2025BTOPermittivity,Li2025PZTMemristor} In contrast, LTOI uses monolithically etched LT waveguides to
achieve ultralow propagation loss, while coplanar-waveguide electrodes can be designed for
velocity-matched high-speed modulation.\cite{Wang2024Nature,Wang2024Optica} The comparisons in Table 1 show that each platform currently
has its own strengths and limitations. The present LTOI platform brings this combination of attributes
into a regime better matched to large-scale high-speed modulation systems, where low-loss routing,
high-speed Pockels modulation and retained phase control are needed in the same material system.

The mechanistic advance of this work is to connect the proposed defect-dipole model with integrated
electro-optic waveguides. Within this model, the limited room-temperature mobility of lithium
vacancies prevents most defect dipoles from fully reorienting during ferroelectric-domain switching.
The built-in field generated by these dipoles therefore retains its original direction and most of
its magnitude after the programming voltage is
removed.\cite{Gopalan1996InternalField,Kim2001DomainReversal} The switched-domain configuration changes
the effective electro-optic coefficient. The retained field interacts with this modified coefficient
to produce a persistent Pockels-induced refractive-index change, converting a stable ferroelectric
state into a retained optical phase state. This mechanism is relevant beyond the
individual devices demonstrated here. Recent thin-film
LN systems, including programmable microwave photonic circuits for RF filtering and signal processing
\cite{Wei2025ProgrammableMWP} and multi-channel WDM transmitter chips \cite{Liu2023LNOIWDM}, show that LN-family photonics is moving towards
circuits that combine many configured optical states with high-speed electro-optic modulation. LT could
serve as a complementary or alternative platform for these applications. By adding non-volatile phase
control, static bias points, resonances and routing states could be written once and then read without
a holding voltage. This would reduce the need for large arrays of thermo-optic tuning elements,
thereby lowering static power, thermal crosstalk, DC control-line count, package pin count, packaging
complexity and the calibration cost associated with long-term drift.\cite{Bogaerts2020Programmable,PerezLopez2020SelfConfiguration,Padmaraju2012Thermal}

The LTOI image-edge-detection chip fabricated in this work illustrates how this platform-level
combination can translate into edge-intelligence hardware. Autonomous vehicles, drones, robots, industrial inspection systems,
smart-security cameras, wearable devices, portable medical instruments and remote-sensing nodes
require local feature extraction under tight constraints on computing power, energy budget and latency,
often in environments with variable sensor noise and interference.\cite{Baek2025EdgeIntelligence} Table 2 captures the resulting
trade-off. Electronic CMOS image-sensor chips are mature and highly integrated, and therefore achieve
low milliwatt-level power consumption, but their reported pixel rates are typically in the
\(\mathrm{Mpixel\,s^{-1}}\) regime.\cite{Jin2020EdgeCIS,Lee2025LowPowerEdgeCIS} Photonic edge-detection processors operate in the
\(\mathrm{Gpixel\,s^{-1}}\) regime.\cite{Xu2022PTFP,Bai2023Microcomb} The present LTOI chip reaches
\(10~\mathrm{Gpixel\,s^{-1}}\) in the current experiment, which is comparable to the representative
photonic processors listed in Table 2. For idealized operation at \(200~\mathrm{Gbit\,s^{-1}}\), the
projected rate reaches \(100~\mathrm{Gpixel\,s^{-1}}\), exceeding the photonic examples compared
there. The deterministic comparison algorithm gives \(100\%\) edge-map
agreement in the noise-free test. This exact criterion setting avoids analogue-weight errors and
provides a larger decision margin against input noise. In terms of energy, electronic chips remain
strong because CMOS integration is compact and mature. Many photonic prototypes, by contrast,
dissipate static power in heater arrays used for WDM stabilization, resonator locking or analogue
weight control.\cite{Xu2022PTFP,Bai2023Microcomb} The LTOI chip removes this on-chip holding power by storing the comparison criteria
non-volatilely. The total system energy nevertheless includes the power required for optical sources,
electrical drivers, photodetection and readout electronics. Further integration and system-level
co-design could reduce these contributions.

Future work should address the present limits of the LTOI platform, including relatively large
programming energy, \(100~\mathrm{ms}\)-scale tuning, a sub-minute stabilization transient and a limited
number of accessible programmed states. These limits are linked to the large coercive field of
congruent LT, depolarization during domain reversal, and charge relaxation and capacitive charging in
the waveguide--electrode structure under large programming fields. Deeper domain engineering, together
with dopant engineering in LT, provides a possible route. For example, Mg-related strategies could be
explored, inspired by the use of Mg-doped periodically poled LN to reduce the poling voltage and
improve domain control.\cite{Hu2003MgSwitching,Ishizuki2003MgPoling} Electrode engineering provides a
complementary route. Transparent conducting oxide (TCO) electrodes could serve as LT non-volatile
programming electrodes. Because they introduce lower optical-mode absorption than metal electrodes,
they can be placed closer to the waveguide, increasing the local poling field for a given voltage and
thereby lowering the required polarization-switching voltage.\cite{Meng2025TCOElectrode} Lower programming voltages enabled by
these material and electrode strategies could also help mitigate charge relaxation and capacitive
transients in the waveguide--electrode structure. In parallel, further optimization of the
programming-pulse parameters could shorten the stabilization time after LT non-volatile tuning. With
continued improvements in programming speed, energy consumption, array uniformity and packaging
electronics, non-volatile LTOI could support large-scale programmable photonic processors, microwave
photonic systems, WDM transmitters, optical interconnects and photonic-computing hardware.\cite{Bogaerts2020Programmable,Wei2025ProgrammableMWP,Liu2023LNOIWDM} Overall,
this work establishes LTOI as an integrated photonics platform that unites low-loss optical routing,
high-speed electro-optic modulation and persistent phase programmability, providing a route towards
scalable photonic circuits with reduced static control overhead.

\section*{Acknowledgements}

This work was supported by the National Natural Science Foundation of China (grant nos. 62205119,
62335014 and T2522012) and the Open Project Program of Wuhan National Laboratory for Optoelectronics (grant no.
2024WNLOKF015). The authors acknowledge the Advanced Micro-Nano Fabrication Center of Huazhong
University of Science and Technology for facility support. Lithium tantalate devices were fabricated
in the cleanroom of Wuhan ANPI Corporation.

\end{document}